\documentstyle[preprint,aps]{revtex}
\begin{document}
\draft
\title{
Pseudogap Formation in the \\
Symmetric Anderson Lattice Model
}
\author{P.G. McQueen \cite{pgmcqueen}, D.W. Hess, }
\address{
Complex Systems Theory Branch, \\ Naval Research Laboratory,
Washington, D.C. \ 20375-5345
}
\author{and J.W. Serene }
\address{ Department of Physics \\
Georgetown University \\
Washington, D.C. \ \ 20057
}
\vspace{0.25in}
\maketitle
\begin{abstract}
We  present self-consistent calculations for the self-energy
and magnetic susceptibility of the 2D and 3D
symmetric Anderson lattice Hamiltonian, in the fluctuation exchange
approximation.   At high temperatures, strong f-electron scattering
leads to broad quasiparticle spectral functions, a reduced
quasiparticle band gap, and a metallic density of states.
As the temperature is lowered, the spectral functions narrow and
a pseudogap forms at the characteristic temperature $T_x$ at which
the width of the quasiparticle spectral function at the gap edge
is comparable to the renormalized activation energy.
 For $T << T_x $,
the pseudogap is approximately equal to the hybridization
gap in the bare band structure.
The opening of the pseudogap is clearly apparent in both the
spin susceptibility and the compressibility.
\\
 \noindent
 PACS numbers: 71.28.+d, 75.20.hr
\end{abstract}
\newpage

It is well known that heavy electron
systems may be superconducting, magnetic or paramagnetic at the
lowest temperatures \cite{fisk88}.  Recent discoveries of cerium
and uranium based compounds with quasiparticle
gaps or pseudogaps $E_g \sim 10-100$K
have generally been interpreted as evidence that the zoo of heavy
electron systems includes semiconductors as well \cite{thompson}.
However, the interpretation of the data is not unambiguous,
and it is not yet clear whether these systems should be
understood simply as ordinary semiconductors with unusually small
gaps and strongly correlated electronic quasiparticles.
An alternate interpretation of the resistivity, susceptibility, specific heat,
and neutron scattering data is that
at high temperatures ($T >> E_g$) these materials most closely
resemble a heavy fermion metal above its coherence temperature, and as the
temperature drops a pseudogap opens in the density of states at
roughly the coherence temperature as estimated from high temperature
properties.

A natural explanation of the existence of the narrow-gap
semiconductors is that the high temperature band structure is
semiconducting, and as the temperature is lowered, the effective
hybridization of the f-levels with the conduction band is reduced
as strong electronic correlations develop, just as in heavy electron metals.
The reduced effective hybridization then leads directly to a dramatic
reduction of the gap from its bare value.  This picture is supported by
slave boson calculations for the Anderson lattice Hamiltonian
with infinite on-site repulsion $U$,
for $T$ below the bose condensation temperature \cite{prise}.
The mean field approach of those calculations does not include the finite
quasiparticle lifetime resulting from strong electron-electron scattering.
We have previously reported calculations of the self-energy
and magnetic susceptibility
for the symmetric Anderson lattice model in 2D with finite $U$,
including the quasiparticle lifetime \cite{lt20}.  From the susceptibility
and DOS we concluded that, on account of the temperature
dependent quasiparticle lifetime, a pseudogap formed with
decreasing temperature at a characteristic temperature
much smaller than the hybridization gap in the bare
band structure.
Jarrell {\em et al.} reported essentially exact
calculations for the  $D = \infty$ symmetric Anderson lattice model with
finite $U$ \cite{jarrell}.
They also observe a metallic high temperature state and the formation
of a pseudogap with decreasing temperature.

Here we present additional results and a detailed analysis of the
formation of the pseudogap in the symmetric Anderson lattice model
with $U \sim W/2$, where $W$ is the bare conduction bandwidth; our results
in 2D and 3D are qualitatively similar.
The Anderson lattice Hamiltonian is \cite{error},
 \begin{eqnarray}
  \! \! \! H  & = &
     - \ t \sum_{<i,j>   \sigma}
  (c_{i \sigma}^\dagger c_{j \sigma } + c_{j \sigma}^\dagger c_{i \sigma })
   + \sum_{i \sigma} \: \{ \; V \, ( f_{ i \sigma}^\dagger
  c_{ i \sigma} + c_{ i \sigma}^\dagger f_{i \sigma} )  \nonumber \\
    &  & + \:
     e_{f} \, n^{f}_{i \sigma}
    - \ h \, \sigma (n^{c}_{i \sigma}
            + n^{f}_{i \sigma} ) \: \}
  +  U  \sum_{i} n^{f}_{i \uparrow} n^{f}_{i \downarrow}\;,
  \end{eqnarray}
where  $c_{i \sigma}$ ($f_{i \sigma}$) annihilates a conduction
(f) electron at site $i$ with spin $\sigma$,
$t$ is the nearest-neighbor conduction electron hopping energy,
$e_f$ is the f-level energy,
$V$ is the onsite hybridization energy between conduction and f electrons,
$U$ is the Coulomb energy of two f-electrons on the same site, and $h$
is a homogeneous magnetic field coupled to the electron spin.  For the
symmetric model $\mu = 0$,  and $e_f$ is chosen so that there are exactly
two electrons per lattice site at all temperatures. The chemical potential
is then at the middle of the indirect hybridization gap in the bare
band structure, between the $\Gamma$ and $M$ points. The bare activation
energy $\Delta_0$ is half the hybridization gap, and given by
\begin{equation}
\Delta_0 = \sqrt{(tD)^2 + V^2} - Dt \, .
\end{equation}

The fully renormalized
single particle dispersion relations determined from the poles of
the retarded Green's function show a temperature
dependent gap $E_g (T)$, which at high temperatures is
close to the bare hybridization gap.  The high-temperature
density of states does not
have a gap, however, because the quasiparticle width at the band
edges, $\Gamma(T)$, is large compared to the temperature dependent
activation energy $\Delta (T) = E_g (T)/2$. In this sense the system
is effectively metallic (though with a poor conductivity).  As the
temperature is lowered, increasing correlations in the metallic state
renormalize $\Delta (T)$ and the gap decreases, but
$\Gamma(T)$ decreases even more rapidly than $\Delta(T)$ until $T$
eventually reaches a crossover temperature $T_x$ where $\Gamma(T_x) \sim
\Delta(T_x)$. At this point the pseudogap begins to open in the single
particle density of states.  The opening of the pseudogap feeds back
into the calculation of the self-energy and leads to
a more rapid decrease of $\Gamma(T)$ and a rapid growth of
$\Delta (T)$; these in turn accelerate the reduction of the density of
states in the pseudogap with decreasing temperature.  At the
lowest temperatures, $\Delta (T) \sim \Delta_0$.
The opening of the pseudogap  is reflected in the temperature
dependence of the spin susceptibility, which drops faster than is
possible for a semiconductor with a fixed gap when $T \lesssim T_x$.
We also note that for $T > T_x$
the self-energy is to an excellent approximation independent of
momentum, but as $T$ is reduced through $T_x$, the self-energy
becomes increasingly momentum dependent.  This is reflected in a
strong flattening of the quasiparticle dispersion at low energy.
With decreasing $U$, $T_x$ increases and the explicit temperature
dependence of the opening of the pseudogap becomes more important;
the temperature dependences of the thermodynamic and single particle
properties are less dramatic and begin to resemble those of ordinary
semiconductors. A crossover temperature $T_x$ can still be identified
from temperature dependence of the self-energy,
and insofar as $T_x(U)$ is meaningful for all $U$, we expect that
for $T_x(U) >> \Delta_0$, $\Delta(T)$ will not differ much from
$\Delta_0$ and the system will behave essentially like an ordinary
insulator.
Here we focus on the more interesting cases $T_x (U) < \Delta_0$.

We have carried out fully self-consistent calculations
for the 2D and 3D symmetric Anderson lattice Hamiltonian using
a self-consistent perturbation theory known as the fluctuation
exchange approximation (FEA) \cite{bsw}.  Here we
present a highly abbreviated description of the FEA as applied
to the Anderson lattice model and refer the reader
to Ref. \cite{alprl} for a more complete description.
In a self-consistent perturbation theory, the fully
renormalized temperature Green's function $G$ is related
to the self-energy $\Sigma$ through Dyson's equation,
\begin{equation}
  G({\bf k}, \varepsilon_n ) = [ G_0^{-1}({\bf k}, \varepsilon_n ) -
 \Sigma({\bf k}, \varepsilon_n )]^{-1}.  \label{dyson}
\end{equation}
The self-energy is obtained from
a generating functional $\Phi [G]$,
\begin{equation}
 \Sigma ({\bf k}, \varepsilon_n ) = \frac{1}{2} \ \frac{ \delta \Phi [G] }
 {\delta G({\bf k}, \varepsilon_n )}, \label{mse}
\end{equation}
and the grand thermodynamic potential is given by \cite{lw}
\begin{equation}
\! \! \! \! \! \!  \Omega(T,\mu, h) = - \ {\rm Tr} \ [ \Sigma G + \ln
   (-G_0^{-1} + \Sigma)] + \Phi [G]\; ,
\end{equation}
which the appealing property of being stationary with respect
to independent variations in $G$ and $\Sigma$.
Here `Tr' denotes a generalized trace over all arguments of the
Green's function and  $G_0$ is the Green's function of the non-interacting
system. With the spin-quantization axis along the field, the
Green's functions and self-energies are spin diagonal, and
because the Hubbard interaction acts only between f-electrons, $\Phi$ is
a functional of the f-electron Green's function $G_{f}$ alone
and only the f-electron self-energy $\Sigma_{f}$ is nonzero.  This
greatly simplifies the solution of Dyson's equation for $G$  \cite{yip}.
The practical application of self-consistent perturbation theory
entails the selection of a specific (possibly infinite) subset of the
diagrams for $\Phi$. The diagrams that define the FEA are
those that generate the Hartree-Fock and second-order self-energies
together with contributions from the exchange of longitudinal spin and
density fluctuations, transverse spin fluctuations, and singlet
pair fluctuations.  Explicit expressions for these contributions
to $\Phi[G]$ for the Anderson lattice model appear in Ref. \cite{alprl}.
For the results presented below, the contribution to $\Phi$ from spin
fluctuations
is within $10 \%$ of that from the second order diagram; contributions from
pair fluctuations and from density fluctuations are smaller by factors of
roughly $-1/5$ and $-1/10$ respectively.

We calculate $\Sigma_{f}$ using an algorithm originally developed for the
Hubbard model \cite{shmb7}, on a $32^{D}$ lattice with spot-checks
on lattices as large as $128 \times 128$ in 2D, and using
a frequency cutoff at least six times the unhybridized
conduction band width, $W = 4 D t$.

The magnetic susceptibility $\chi_s(T)$ was found from the slope of
${\rm Tr}[\sigma_{z} G]$ versus $h$ for small fields
(typically $0.005t$ and $0.01t$); this calculation of $\chi_s(T)$
is equivalent to solving an integral equation for the response
function with irreducible vertices comprising particle-particle
and particle-hole bubble-chains and Aslamazov-Larkin-type
diagrams \cite{al}.

The retarded self-energy on the real axis was obtained from
N-point Pad\'{e} approximants \cite{vs}.  This self-energy was then used
to construct the spectral functions and quasiparticle dispersion
relations along symmetry directions in the Brillouin zone.
The densities of states were found by analytic continuation
of the ${\bf k}$-summed imaginary-frequency Green's functions
using a varient of the N-point Pad\'{e} approximant method of Vidberg and
Serene \cite{lr}.
These densities of states satisfy sum rules to better than $ 1\%$ accuracy.
At some temperatures in the transition region
between the metallic and insulating states (see below), it
is difficult to obtain reliable analytic continuations.
For example, Fig.~1 compares three different Pad\'{e} continuations
for the DOS at two temperatures: $T=0.055t$,
where the gap is opening, and $0.035t$
where the gap is well formed.  The three analytic continuations for $T =
0.035t$
are essentially indistinguishable for all frequencies.  At low energy, the
analytic
continuations for $T=0.055t$ agree well except very near $\varepsilon = 0$;
for $2t \leq \varepsilon \leq 8t$, two of the three continuations nearly
coincide while the third shows qualitatively different behavior.  To
investigate this variability,  we computed the rms deviation of the
original imaginary frequency data from the data generated by using the
candidate analytic continuation as a spectral function.
For this purpose we use only frequencies less than
half the high frequency cutoff to avoid cutoff-dependent artifacts.
For $T = 0.035t$, the curves shown have rms deviations of $1.3 \times 10^{-4}$,
$1.1 \times 10^{-4}$ and $1.3 \times 10^{-4}$ for $N = 57$, $59$, and $61$
respectively.
At the temperature where the pseudogap is opening, qualitative differences
in the structure of the DOS at energies $ \sim W/2$ are often weakly reflected
in the rms deviation as seen for the $T = 0.055t$ DOS shown in Fig.~1. There
the rms deviations are $5.4 \times 10^{-5}$, $5.4 \times 10^{-5}$
and $6.1  \times 10^{-5}$ for $N = 35$, $37$, and $39$ respectively;
the qualitatively different continuation shows only a slightly larger
rms deviation.  The DOS results presented below will include
only those temperatures at which the analytic continuations for a
given (reasonable) $N$ agree well with those over a range of $N$
and have the smallest rms deviations. In all cases
the rms deviation is less than $2 \times 10^{-4}$.

For the calculations in 2D reported here, we used
$\mu = 0$, $V = t$ and $\tilde{e}_f = 0$, where
$\tilde{e}_f = e_f + \case 1/2 U(n_{\uparrow}^f + n_{\downarrow}^f)$
is the ($h = 0$) Hartree-renormalized f-level.  With this choice of
parameters the bare activation energy $\Delta_0$ is $0.236t$.
For the 3D model, we took $V = 1.5 t$
so that the ratio of the bare hybridization gap to the conduction
electron bandwidth is the same in 3D as in 2D.
We found that in this case the spin susceptibility and
density of states for the symmetric Anderson lattice model
in 3D may be roughly scaled to those in 2D by scaling
all energies with respect to the bandwidth. An example of this
scaling for the spin susceptibilities
$\chi_s$ for $U=W/2$ is shown in Fig.~2. In the following
we present primarily 2D data.

As the temperature is decreased through $T_x$, a pseudogap opens in
the single particle
excitation spectrum.  Pseudogap formation is reflected in $\chi_s$
and in the density $n(\mu, T)$; analytic continuation is not needed
to calculate either of these. Fig.~3a shows $\chi_s(T)$ for various
$U$ including $U=0$.  The form of $\chi_s(T)$ for finite $U$
strongly suggests the existence of a gap that apparently decreases
with increasing $U$. For the largest $U$ shown,
$U=4t = W/2$, the rapid decrease in $\chi_s(T)$ occurs at a temperature
$\sim \Delta_0 / 4$.
For $U > 0$, $\chi_s(T)$ cannot be scaled to the $U=0$ result
by a simple renormalization of the gap.  The dependence of $n$ on
$\mu$ also
shows a clear signature of pseudogap formation, as shown in Fig.~3b.
For $U=0$ and  $T \gtrsim \Delta_0$, $d n/ d \mu$ is constant
over $-0.3t \lesssim \mu \lesssim 0.3t$ whereas for
$T = 0.0625 t$, $d n/ d \mu$ shows substantial curvature.
As the temperature is reduced, the gap in the single particle excitation
spectrum becomes more apparent and $d n/ d \mu$ essentially vanishes for
$\mu \lesssim \Delta_0$.  In contrast, for $U = W/2$
and $T = 0.0625 t$, $d n/ d \mu$ is essentially constant
for $-0.3 < \mu < 0.3$, consistent with a metallic state or a semiconducting
state with a gap $\lesssim 0.0625 t$.  At the lower temperature of
$T = 0.055t$, $n(\mu)$ shows slight curvature.
As the temperature is lowered, $d n/ d \mu \rightarrow 0$ for a
region of $\mu$ about $\mu = 0$ as expected in the presence of
a pseudogap.  At the lowest temperature, the region where $d n/ d \mu$
nearly vanishes is about the size of the bare hybridization gap.
{}From $n(\mu)$ and $\chi_s(T)$ one estimates essentially the same
characteristic
temperature $T_x$ for pseudogap formation.

The formation of the pseudogap can be seen explicitly in the
single-particle (tunneling) density of states shown for various
temperatures in Fig.~4. For $T=0.125t$ and $T=0.03125 t$, Fig.~4a
shows the total DOS over a wide energy range,
from which one sees that the density of states is smoothly varying
over most of the bandwidth.
For $T=0.125t$ there is no sign of a gap at low energy, but for
$T = 0.03125$ a pseudogap centered at $\varepsilon=0$ is clearly
evident.  Over the narrower range $-t \leq \varepsilon \leq t$ shown
in Fig.~4b there is still no evidence of a hybridization gap at the
highest temperatures, but as the temperature is lowered, a sharp
asymmetric peak signals the pile up of states
at the pseudogap edge.  The density of states at zero energy decreases
rapidly with decreasing temperature.  We have shown previously that this
decrease closely tracks that of $\chi_s(T)$ \cite{lt20,note2}.

To illustrate how the pseudogap forms, we plot in Fig.~5a
the spectral functions as a function of $\varepsilon$ at the $\Gamma$ and
$M$ points which lie on either side of the pseudogap.
At high temperature, the spectral functions are very broad due to
strong
electron-electron scattering. The tails of the two spectral functions,
centered
close to the bare gap edges, extend well into
the gap where they overlap and form the observed maximum.  As the
temperature is
decreased, the spectral functions at the $\Gamma$ and $M$ points
sharpen as shown in Fig.~5b.
In contrast to the relatively strong $T$ dependence of the
width of the spectral function $\Gamma (T)$, the position of the
quasiparticle band at the $\Gamma$ point $\Delta(T)$ (as determined from the
peak of the spectral function), also shown in Fig.~5b,
is weakly $T$ dependent until $\Gamma (T) \sim \Delta(T)$, which
sets the crossover temperature $T_x$. At this
point the dramatic rapid decrease in $\chi_s (T)$ and the low
energy DOS begins, with
a correspondingly rapid renormalization of the quasiparticle energy
at the gap edge.

The correlations above $T_x$ are apparent in the large renormalization of
the band structure at low energy; the peak positions of the spectral
functions at the gap edges at $T_x$ are at about $\Delta_0 / 4$.
The sharpening of the spectral functions with decreasing temperature above
$T_x$
reflects the evolution toward coherent quasiparticle excitations.
The quasiparticle renormalization
factor provides another measure of the strength of correlations.
In a metal,  the quantity
${\rm Im} \, \Sigma({\bf k}, \varepsilon_0)/\varepsilon_0 = {\cal D}_{k}(T)$
can be used to construct a reasonable estimate of the quasiparticle
renormalization
factor $a_k = ( 1 - \partial \, {\rm Re} \, \Sigma({\bf k},\varepsilon)/
\partial \varepsilon |_{\varepsilon = 0} )^{-1}$;
for a point on the Fermi surface of a Fermi liquid, ${\cal D}_{k_F}(T)$ tends
to
$\partial \, {\rm Re} \, \Sigma(\vec{k}_F, \varepsilon) / \partial \varepsilon
|_{\varepsilon =0}$
as $T \rightarrow 0$ \cite{sh}.  For the one band Hubbard model at $1/4$
filling, we
found that ${\cal D}_{k_F}(T)$ is weakly temperature dependent and approaches
$\partial \, {\rm Re} \, \Sigma({\bf k},\varepsilon)/ \partial \varepsilon
|_{\varepsilon = 0}$
\cite{shmb7}.
For the asymmetric Anderson lattice model \cite{alprl}, we found that
the magnitude of ${\cal D}_k(T) $ increases strongly with
decreasing temperature (as did our measure of the quasiparticle effective mass)
down
to the lowest temperatures that we studied.  Here we observe that for $T > T_x$
the
temperature dependence of ${\cal D}_k (T) $ resembles that of ${\cal
D}_{k_F}(T) $ for the
asymmetric model.  We plot ${\cal D}_k (T) $ for the symmetric model and for
${\bf k}$ at
the $\Gamma$ point in Fig.~6, in which a distinctive feature at $T = T_x$ is
apparent. For $T > T_x$, the system is metallic
and the magnitude of ${\cal D}_\Gamma (T)$ increases with
decreasing temperature.  As the
temperature is further decreased, the opening of the pseudogap acts to
reduce contributions from low energy excitations to the fluctuation
propagators, leading to a sharp reduction in the magnitude of
${\cal D}_\Gamma (T)$, which is reflected in the
widening of the renormalized gap shown in Fig.~5b.  In the limit $T \rightarrow
0$,
${\rm Im} \, \Sigma({\bf k}, i \varepsilon_n)$ tends to a finite value at low
$T$.
This is not inconsistent with the formation of the pseudogap;
the $U=0$ Green's function already contains
information on the bare hybridization gap, so
$\Sigma({\bf k}, \varepsilon_0)$
need not show a divergence as observed, e.g. in the Hubbard model \cite{white}.

Effects of strong correlations are also apparent in the
quasiparticle band structure determined from the poles of the single particle
Green's function,  as shown in Fig.~7.  At sufficiently high temperatures, the
quasiparticle band structure does not differ significantly from the bare band
structure and $\Sigma({\bf k}, \varepsilon)$ is to a good approximation
independent
of ${\bf k}$.  As the temperature is lowered toward $T_x$, a pronounced
flattening
of the bands around the $\Gamma$ and $M$ points occurs and the self-energy
becomes
increasingly ${\bf k}$ dependent.  In Fig.~7b we compare the band structure
obtained by taking $\Sigma({\bf k}, \varepsilon) =\Sigma( \Gamma, \varepsilon)$
for all ${\bf k}$ with that obtained from the full ${\bf k}$-dependent
self-energy.
The momentum dependence of the self energy clearly provides a substantial
contribution
to the flattening of the band at low energy.  Mean field
calculations in the limit $U \rightarrow \infty$ find a quasiparticle
band structure given by the expression for the bare band structure
with a renormalized f-level and hybridization.
In Fig.~7b we also show the band structure for a noninteracting system
along the $\Gamma-X$ direction for
$e_f = 0$ and $V = 0.6t$, which reproduces the particle-hole symmetry and gap
of the fully renormalized band structure.  It is clear that this approach is
unable to describe the band flattening and band width along this
direction in the FEA.

We have presented the results of a fully self-consistent calculation of
the self-energy and spin susceptibility in the fluctuation exchange
approximation for the symmetric Anderson lattice model.
{}From a metallic state at high temperature, a pseudogap forms
with decreasing temperature.  The pseudogap begins to open
at the temperature $T_x$ at which the renormalized band gap
and the width of the spectral function at the gap edge are comparable.
As $T$ is decreased through $T_x$, the density of states at low energy
decreases rapidly. The opening of the pseudogap
is observed in results obtained directly from imaginary
frequency axis data and through analytic continuation to
the real frequency axis.   The large temperature-dependent
renormalizations of the band structure above $T_x$ are similar to
those we have observed in the asymmetric model and suggest
that the evolution of a coherent heavy
fermion metallic state with decreasing $T$ is cut off by the
loss of low energy excitations as the pseudogap develops.

\acknowledgments

It is a pleasure to acknowledge useful discussions with M. Jarrell,
J.D. Thompson and M.F. Hundley. We are grateful to J. Deisz for
a critical reading of this manuscript.
We thank the Office of Naval Research for their support and
the NRL Connection Machine facility for assistance.
This work was supported in part by the Office of Naval Research and by
the Army High Performance Computing Center under the auspices of the Army
Research Office contract number DAALO3-89-C-0038 with the University of
Minnesota.

\begin{figure}
\caption{
 Comparison of three analytic continuations using N-point Pad\'{e} approximants
 for $T = 0.055t$ and $0.035t$ (a) low energies, $0 \leq \varepsilon
 \leq t$;  and (b) high energies, $2t \leq \varepsilon \leq 8t$.
 For $T = 0.055t$ ($0.035t$),
 35, 37 and 39 (57, 59 and 61) Pad\'{e} coefficients were used.  Note that the
 ``high energy'' $T = 0.035t$ results were shifted by $0.05$ for clarity and
that
  35 coefficients were used for the dashed $T = 0.055t$ curve.}
\label{fig1}
\end{figure}

\begin{figure}
\caption{The uniform static spin susceptibility $\chi_s$ for 2D
 and $U = 4t = W/2$  ($\bullet$), 3D and $U = 6t = W/2$ ($\triangle$), and the
3D result
 with all energies scaled by the ratio of the bare conduction bandwidth in 2D
to that in
 3D bar($\circ$).
}
\label{fig2}
\end{figure}

\begin{figure}
\caption{(a) The spin susceptibility $\chi_s (T)$ for $U = 0,$ (dashed) $ t
(\Box)$, $2t (\circ)$, $3t (+)$ and $4t (\bullet)$. (b) The total density $n$
as a function of $\mu$ for $U = 4t$ and $T = 0.0624t (\triangle), \ 0.055t
(+), \ 0.03125t (\diamond),$ and $0.015625 (\circ)$. Also shown for comparison
are $n (\mu)$ for $U=0$ and $T = 0.24t$ (solid), $0.0625t$ (long dashed),
and $0.015625t$ (short dashed).
}
\label{fig3}
\end{figure}

\begin{figure}
\caption{
(a) The density of states over a wide energy range for $T = 0.125t$
(short-dashed) in the metallic state and for $T=0.035t$ (long-dashed) where
a developed pseudogap is evident.
(b) The density of states at low energy for temperatures $T= 0.015625t$
(long-dashed showing gap), $0.035t$ (short-dashed showing gap), $0.0625t$
(long-dashed-short-dashed),
$0.125t$ (dashed), and $0.25$ (dotted) showing the formation of a pseudogap
from
a high temperature metallic state
 with decreasing $T$.
}
\label{fig4}
\end{figure}

\begin{figure}
\caption{(a) The spectral function at the $\Gamma$ point (solid) for $T= 0.25t,
\
 0.125t, \ 0.0625t$, and $0.035t$ in order of decreasing peak width.  Also
shown is
 the spectral weight for the $M$ point (dashed) and the sum of the $M$ and
$\Gamma$
 point spectral weights (long dash) for the temperature $T = 0.0625t$ at which
the
 pseudogap begins to open for $U = 4t$. (b) The width of the spectral function
 at half maximum on the high energy side $\triangle$ and low energy side
$\circ$
 as a function of $T$ for ${\bf k}$ at the zone center, and the position of
 the peak of the spectral function ($\bullet$) as a function of temperature.
 }
\label{fig5}
\end{figure}

\begin{figure}
\caption{The function ${\cal D}_{\bf k}(T)$ (see text) for ${\bf k}$ at the
$\Gamma$
  point and for $U = t \ (\circ), \ 2t \ (\Box), \ 3t \ (\diamond),$
  and $4t \ (\bullet)$. Note the sharp structure marking
  the opening of the pseudogap and the shift of this structure to lower
  temperature with increasing $U$.}
\label{fig6}
\end{figure}

\begin{figure}
\caption{(a) Fully renormalized quasiparticle band structure along high
symmetry
  directions in the zone for $U=4t$ with $T=0.0625t$ (solid) and $T= 0.25$
(dashed)
  compared with the bare band structure (long dash-short dash).
  (b) Fully renormalized band structure for $T=0.0625t$ (solid) compared with
   the band structure assuming a momentum independent self energy
   $\Sigma ({\bf k}, \varepsilon) = \Sigma ( \Gamma, \varepsilon)$ (dashed),
the
   bare band structure (short-dashed) and a fit to the
   mean field band structure with
   a renormalized $V = 0.6t$ and $e_f = 0$ (long dash-short dash).
}
\label{fig7}
\end{figure}


\begin{references}
\bibitem{error} Note that an unfortunate error in the hopping term
   of the Hamiltonian in Eq. [1] of Ref. \cite{alprl} is corrected here
   so that the Hamiltonian in Eq. [1] is consistent with the notation
   used in our previous papers.
\end{references}
\end{document}